\documentclass[sigconf]{acmart}

\usepackage{booktabs}
\usepackage{tikz}
\usetikzlibrary{arrows.meta, positioning, shapes.geometric}
\AtBeginDocument{%
  }

\setcopyright{none}
\begin{document}

\title{From Financial Sentiment Classification to Return Predictability:
A QLoRA Benchmark of Large Language Models}

\author{Fusheng Luo}
\authornote{This work was conducted while the author was a graduate student at Johns Hopkins University.}
\email{fluo5@alumni.jh.edu}
\orcid{0009-0000-8099-9835}
\affiliation{%
  \institution{Johns Hopkins University}
  \city{Baltimore}
  \state{Maryland}
  \country{USA}
}

\renewcommand{\shortauthors}{Luo}

\begin{abstract}
Financial sentiment classifiers are commonly evaluated against human labels,
but strong linguistic performance does not necessarily imply economically
useful return predictability. This study separates these questions through two
experiments. First, we construct a unified three-class benchmark from five
financial text datasets and compare TF--IDF Naive Bayes, off-the-shelf
FinBERT and Financial-RoBERTa encoders, zero-shot Qwen2.5-7B, and
QLoRA-adapted Qwen2.5-7B, LLaMA3-8B, and Mistral-7B models. Mistral-7B
achieves the best test accuracy (0.8840) and macro-F1 (0.8771), while QLoRA
raises Qwen2.5's macro-F1 from 0.7274 to 0.8615. An inverse-frequency
class-weighted loss does not improve Qwen2.5. Second, we evaluate economic
validity on a temporally separate 2019 Benzinga sample containing 10,637 unique
headlines and 13,115 headline--stock observations for a fixed S\&P~100
universe. Model probabilities are converted into continuous sentiment scores,
aggregated by stock and signal date, and aligned with next-session returns over
one-, two-, three-, and five-day horizons. All seven downstream models produce
positive but small mean rank information coefficients at the one-day horizon;
the largest is 0.0143 for FinBERT. None of the 28 model--horizon tests remains
significant after Newey--West inference and false-discovery-rate correction.
Portfolio results likewise fail to establish a robust advantage for the
best-performing classifiers. The findings show that QLoRA is effective for
financial sentiment adaptation, while also documenting a clear gap between
classification accuracy and tradable cross-sectional signals.
\end{abstract}


\maketitle

\section{Introduction}

Financial markets continuously absorb information from news, policy
communications, corporate disclosures, analyst commentary, and social media.
Automatically identifying the polarity of these texts is therefore useful for
information retrieval, risk monitoring, and quantitative research. The task is
nevertheless more demanding than general-domain sentiment analysis: financial
polarity depends on expectations, entity context, and market interpretation,
and neutral language is common. Domain-specific encoders such as FinBERT have
addressed part of this problem, while recent instruction-tuned large language
models (LLMs) offer broader contextual capacity at substantially greater
computational cost \cite{araci2019finbertfinancialsentimentanalysis}.

Parameter-efficient fine-tuning (PEFT) provides a practical way to adapt these
models. In particular, QLoRA combines a frozen 4-bit quantized backbone with
trainable low-rank adapters, making task-specific optimization of billion-scale
models feasible on limited hardware \cite{hu2021loralowrankadaptationlarge,
dettmers2023qloraefficientfinetuningquantized}. Existing research has shown
strong results for PEFT in financial NLP, but comparisons are often conducted
on individual datasets or report only linguistic metrics. Two questions thus
remain intertwined: whether QLoRA reliably improves financial sentiment
classification across heterogeneous sources, and whether a more accurate
classifier produces a more informative market signal.

We address these questions through two deliberately separated experiments.
Experiment~1 evaluates sentiment classification on a harmonized benchmark of
five datasets using a fixed train--validation--test protocol. It compares a
traditional TF--IDF Naive Bayes baseline, two off-the-shelf domain-specific
encoders, zero-shot Qwen2.5, three QLoRA-adapted LLM backbones, and a
class-weighted-loss ablation. Experiment~2 then applies seven fixed
probability-producing classifiers---the traditional baseline, two
domain-specific encoders, and four QLoRA specifications---to an unlabeled,
temporally separate 2019 Benzinga sample for a fixed S\&P~100 universe.
Continuous model probabilities are evaluated using daily cross-sectional rank
information coefficients and equal-weighted portfolios over one-, two-, three-,
and five-day horizons.

The study makes three contributions. First, it provides a unified three-class
comparison across heterogeneous financial text sources and multiple 7--8B LLM
families. Second, it isolates the contribution of QLoRA through a same-backbone
zero-shot comparison and tests whether inverse-frequency loss weighting helps
under moderate class imbalance. Third, it evaluates downstream economic
validity with horizon-aware return alignment, Newey--West inference, and
false-discovery-rate correction. The results show substantial gains from QLoRA
on labeled sentiment data, but no statistically robust return predictability,
highlighting that semantic accuracy and economic usefulness are distinct
objectives.

The remainder of the paper reviews related work, defines the two experimental
protocols, reports the classification and downstream results, and concludes
with limitations and directions for intraday, risk-adjusted evaluation.

\section{Background and Related Work}

\subsection{Financial Sentiment Analysis}
Financial sentiment analysis aims to automatically identify market-related sentiment and opinions expressed in financial texts, including news articles, earnings reports, analyst reports, corporate disclosures, central bank communications, and investor discussions. As financial markets generate vast amounts of unstructured textual information every day, accurately extracting sentiment from these sources has become an important task in financial natural language processing (NLP). Unlike traditional structured financial indicators, textual information often contains timely signals regarding corporate performance, macroeconomic conditions, policy expectations, and investor confidence, all of which may influence market behavior and investment decisions.

Despite sharing the same objective of sentiment prediction as general-domain sentiment analysis, financial sentiment analysis presents unique challenges due to the specialized nature of financial language. In financial contexts, sentiment is often associated with market expectations and future economic implications rather than explicit emotional expressions. Consequently, correctly interpreting financial texts requires not only linguistic understanding but also domain-specific knowledge and contextual reasoning. These characteristics have motivated extensive research into financial sentiment analysis over the past decade, leading to the development of specialized benchmark datasets, domain-adapted language models, and increasingly sophisticated learning paradigms for financial NLP.

Unlike general-domain sentiment analysis, financial sentiment analysis requires models to interpret language from the perspective of financial markets rather than human emotions alone. In many cases, the sentiment conveyed by a financial statement depends not only on the lexical polarity of individual words but also on the underlying economic context and market expectations. This distinction was first systematically highlighted by Loughran and McDonald, who demonstrated that general-purpose sentiment dictionaries substantially misclassify financial terminology. Words such as \textit{liability}, \textit{capital}, and \textit{tax} are frequently labeled as negative in general lexicons despite carrying neutral or technical meanings in corporate disclosures, motivating the development of finance-specific sentiment resources \cite{loughran2011liability}.

Beyond domain-specific vocabulary, financial sentiment is inherently context-dependent. Identical expressions may imply opposite market signals depending on surrounding information. For example, reporting a ``loss’’ is not necessarily interpreted negatively if the result exceeds market expectations, while seemingly positive financial results may trigger adverse market reactions when they fall short of investors’ forecasts. Consequently, financial sentiment is more appropriately viewed as a reflection of expected market interpretation than explicit emotional polarity. This market-oriented nature has driven the evolution of financial sentiment analysis from simple lexicon-based approaches toward context-aware language models capable of capturing subtle semantic relationships and financial reasoning, ultimately motivating the development of increasingly diverse benchmark datasets for evaluating financial NLP systems.

To facilitate the development and systematic evaluation of financial sentiment analysis, a series of benchmark datasets has been introduced over the past decade, reflecting the increasing diversity of financial text sources and research objectives. Representative benchmarks are summarized in Table~\ref{tab:financial_datasets}. Among them, Financial PhraseBank has become the de facto benchmark for sentence-level financial sentiment classification due to its expert annotations and carefully curated financial news sentences, establishing a standardized setting for evaluating supervised sentiment classification models \cite{malo2014GoodDO}.

As research progressed, benchmark datasets gradually expanded beyond sentence-level news classification toward more realistic financial scenarios. The FiQA benchmark introduced aspect-level financial opinion mining by incorporating both financial news and microblog posts, encouraging models to identify sentiment toward specific financial aspects rather than assigning a single sentiment label to an entire sentence \cite{fiqa}. Social-media-oriented datasets, such as Twitter Financial News Sentiment, further broadened the scope by providing short, rapidly evolving financial texts that better capture real-time market information and investor attention, while also introducing greater linguistic variability and noise. These datasets have become increasingly valuable for evaluating modern transformer models and large language models under realistic market conditions.

More recent benchmarks have extended financial sentiment analysis beyond traditional news sentiment classification. Datasets constructed from Federal Open Market Committee (FOMC) statements focus on monetary policy communications, where sentiment is expressed implicitly through policy guidance rather than explicit emotional language \cite{shah-etal-2023-trillion}. Such datasets require models to understand nuanced economic reasoning and macroeconomic context instead of relying solely on sentiment-bearing words. At a finer level of granularity, SEntFiN 1.0 introduced entity-aware sentiment annotations, allowing multiple financial entities within the same news headline to receive different sentiment labels \cite{Sinha_2022}. This setting more closely reflects real-world financial news, in which a single document may simultaneously convey positive information about one company while expressing negative sentiment toward another.

Overall, the evolution of benchmark datasets illustrates a clear transition in financial sentiment analysis—from sentence-level polarity classification to aspect-aware, policy-oriented, and entity-level financial language understanding. Rather than simply increasing dataset size, recent benchmarks have progressively emphasized richer financial contexts, finer annotation granularity, and more realistic application scenarios, thereby providing increasingly challenging evaluation environments for modern language models.

Although the availability of benchmark datasets has substantially advanced financial sentiment analysis, the increasing diversity of these resources also introduces new challenges for model evaluation and comparison. Existing benchmarks differ considerably in text sources, annotation schemes, sentiment granularity, and downstream objectives. For example, datasets may focus on financial news, social media, monetary policy communications, or entity-specific sentiment, each reflecting different linguistic characteristics and requiring different levels of financial reasoning. As a result, models that perform well on one benchmark may not necessarily generalize to other financial text domains, highlighting the importance of evaluating robustness under domain shift rather than relying solely on in-domain performance \cite{taxonomy_nlp}.

Another challenge arises from the lack of a unified evaluation protocol across financial sentiment benchmarks. Existing studies employ different sentiment taxonomies, ranging from sentence-level polarity classification to aspect-based and entity-level sentiment analysis, making direct comparison across datasets difficult. Furthermore, many financial sentiment datasets exhibit substantial class imbalance, with neutral samples often dominating the data distribution. Consequently, evaluation metrics such as macro-F1 have become increasingly important because they provide a more balanced assessment of model performance across sentiment classes, complementing conventional accuracy measures \cite{entity_classification}.

Taken together, these observations indicate that the next stage of financial sentiment analysis is no longer driven solely by the construction of larger benchmark datasets, but by the development of evaluation frameworks capable of assessing model robustness across heterogeneous financial scenarios. Such a perspective is particularly relevant in the era of large language models, where comparing traditional machine learning approaches, domain-specific pretrained language models, and parameter-efficient fine-tuned LLMs under a unified experimental setting has become increasingly important for understanding their practical applicability in financial sentiment classification.

\begin{table*}[t]
\caption{Representative benchmark datasets that have shaped the development of financial sentiment analysis.}
\label{tab:financial_datasets}
\centering
\small
\begin{tabular}{p{3.0cm}p{0.8cm}p{2.6cm}p{2.3cm}p{7.3cm}}
\toprule

\textbf{Dataset} &
\textbf{Year} &
\textbf{Source} &
\textbf{Task} &
\textbf{Research Focus} \\

\midrule

Financial PhraseBank
&
2014
&
Financial News
&
Sentence Classification
&
Expert-annotated benchmark for sentence-level financial sentiment classification. It remains one of the most widely adopted benchmark datasets for evaluating financial sentiment models.
\\

FiQA
&
2018
&
News \& Microblogs
&
Aspect-based Sentiment Analysis
&
Introduces fine-grained financial opinion mining and aspect-level sentiment analysis using both financial news and social media data.
\\

Twitter Financial News Sentiment
&
--
&
Financial Tweets
&
Tweet Classification
&
Provides short financial social-media posts labeled as bearish, neutral, or bullish, adding timely but linguistically noisy market commentary.
\\

FOMC Statements
&
2023
&
Monetary Policy Documents
&
Policy-Stance Classification
&
Classifies hawkish, neutral, and dovish policy language, requiring models to capture implicit monetary-policy signals and forward guidance.
\\

SEntFiN 1.0
&
2022
&
Financial News
&
Entity-level Sentiment Analysis
&
Introduces multi-entity sentiment annotations, enabling fine-grained sentiment prediction for multiple financial entities within the same document.
\\

\bottomrule
\end{tabular}
\end{table*}

\subsection{Evolution of Financial Sentiment Analysis Methods}
Early research on financial sentiment analysis primarily relied on lexicon-based approaches, which estimate sentiment by aggregating the polarity of sentiment-bearing words within a document. These methods were initially adapted from general-domain sentiment analysis using manually constructed sentiment dictionaries. However, researchers soon recognized that general-purpose lexicons perform poorly on financial texts because many financial terms convey technical rather than emotional meanings. This limitation motivated the development of domain-specific financial lexicons, among which the Loughran–McDonald (LM) Financial Sentiment Dictionary has become the most influential. By redefining sentiment categories according to financial reporting practices, the LM dictionary substantially improved sentiment estimation in corporate disclosures and established a widely adopted baseline for financial text analysis \cite{karanikola2023financial}.

Despite their simplicity and strong interpretability, lexicon-based methods rely heavily on predefined vocabularies and treat sentiment as the aggregation of individual word polarities. Consequently, they struggle to capture contextual semantics, negation, compositional meaning, and implicit market expectations. As financial language became increasingly diverse and benchmark datasets expanded beyond corporate reports to news, social media, and policy communications, purely dictionary-based approaches gradually reached their performance limits. These limitations motivated the transition toward supervised machine learning methods capable of learning sentiment representations directly from labeled financial text \cite{karanikola2023financial}.

The emergence of deep learning fundamentally transformed financial sentiment analysis by shifting the focus from manually engineered features to automatic representation learning. Early neural approaches, including convolutional neural networks (CNNs) and recurrent neural networks (RNNs) such as Long Short-Term Memory (LSTM) networks, enabled models to learn distributed semantic representations directly from financial text. Compared with sparse representations such as Bag-of-Words and TF–IDF, neural models captured richer syntactic and semantic information, improving their ability to recognize contextual relationships and long-range linguistic dependencies. Nevertheless, sequential neural architectures remained limited in modeling complex contextual interactions over long financial documents and often required substantial task-specific training data to achieve competitive performance.

The introduction of transformer-based pretrained language models marked a major milestone in financial natural language processing. By leveraging large-scale self-supervised pretraining followed by task-specific fine-tuning, models such as BERT demonstrated remarkable improvements across a wide range of NLP tasks through their ability to learn contextualized bidirectional language representations. Among transformer-based models, FinBERT \cite{araci2019finbertfinancialsentimentanalysis} has become one of the most influential baselines for financial sentiment analysis. FinBERT introduces a BERT-based language model specifically adapted to the financial domain through domain-adaptive pretraining on large-scale financial corpora, followed by task-specific fine-tuning for downstream financial NLP tasks. Experimental results demonstrate that domain-specific pretraining substantially improves financial sentiment classification over general-purpose BERT models, establishing FinBERT as a strong benchmark for financial sentiment analysis and a widely adopted baseline in subsequent studies.

Although transformer-based models significantly advanced financial sentiment analysis by providing stronger contextual understanding and domain-specific language representations, they were primarily designed as encoder-only architectures optimized for discriminative tasks such as classification. The rapid emergence of large language models has subsequently shifted research attention toward instruction-following, generative reasoning, and cross-task generalization, raising a new question of whether foundation models can further improve financial sentiment understanding beyond specialized encoder-based models such as FinBERT. This transition has motivated the next generation of research on financial large language models and parameter-efficient adaptation techniques.

\subsection{Large Language Models for Financial NLP}
The emergence of large language models (LLMs) has fundamentally reshaped natural language processing by extending pretrained language models beyond task-specific representation learning toward instruction following, generative reasoning, and cross-task generalization. Unlike encoder-based models such as BERT and FinBERT, decoder-only and encoder–decoder foundation models are pretrained on massive text corpora using self-supervised objectives and subsequently adapted to a wide variety of downstream tasks through prompting or lightweight fine-tuning. Representative open-source models, including LLaMA, Mistral, and Qwen, have demonstrated remarkable performance across diverse NLP applications while significantly lowering the barrier for domain adaptation through publicly available model weights and instruction-tuning techniques.

These advances have substantially influenced financial NLP research. Instead of developing separate architectures for individual financial tasks, researchers increasingly investigate whether a single foundation model can support a broad spectrum of applications, including financial sentiment analysis, question answering, information extraction, report summarization, forecasting, and financial reasoning. Compared with earlier transformer-based classifiers, LLMs provide stronger contextual understanding, superior zero-shot and few-shot capabilities, and greater flexibility for adapting to heterogeneous financial tasks \cite{jadhav2025large}. Consequently, recent research has shifted from designing task-specific architectures toward adapting general-purpose foundation models to the financial domain through domain-specific pretraining, instruction tuning, and parameter-efficient fine-tuning .

The rapid adoption of large language models has led to the emergence of a comprehensive financial LLM ecosystem, encompassing specialized foundation models, open-source adaptation frameworks, and standardized evaluation benchmarks. One of the earliest milestones was BloombergGPT, a 50-billion-parameter language model pretrained on a large mixture of proprietary financial data and general-domain corpora \cite{wu2023bloomberggptlargelanguagemodel}. Unlike domain-adaptive models such as FinBERT \cite{araci2019finbertfinancialsentimentanalysis}, BloombergGPT demonstrated that training a foundation model specifically for finance could substantially improve performance across a wide range of financial NLP tasks while maintaining competitive general-language capabilities. However, its proprietary training data and closed-source nature limited reproducibility and broader research adoption.

To improve the accessibility of financial LLM research, FinGPT proposed an open-source and data-centric framework that leverages publicly available financial data together with parameter-efficient fine-tuning techniques, including LoRA and QLoRA, to adapt general-purpose LLMs for financial applications \cite{yang2025fingptopensourcefinanciallarge, hu2021loralowrankadaptationlarge, dettmers2023qloraefficientfinetuningquantized}. Rather than training a new foundation model from scratch, FinGPT demonstrated that competitive financial language models could be obtained through efficient adaptation pipelines at a fraction of the computational cost required by proprietary systems \cite{yang2025fingptopensourcefinanciallarge}. This open-source paradigm has significantly lowered the barrier for developing financial LLMs and accelerated research on downstream tasks such as sentiment analysis, financial question answering, information extraction, and market forecasting. 

As financial LLMs continue to evolve, standardized evaluation has become increasingly important. FinBen addresses this challenge by introducing a comprehensive benchmark covering dozens of datasets across a broad spectrum of financial tasks, including textual analysis, question answering, information extraction, forecasting, decision-making, and retrieval-augmented applications \cite{xie2024finben}. Its large-scale empirical evaluation demonstrates that while modern LLMs exhibit strong performance on information extraction and conventional text classification tasks, they continue to face substantial challenges in complex financial reasoning and decision-oriented applications. Collectively, these studies indicate that recent research has shifted beyond constructing increasingly larger financial language models toward developing open, efficient, and systematically evaluated financial AI systems.

\subsection{Parameter-Efficient Adaptation of Large Language Models}
Although large language models have demonstrated remarkable capabilities across financial NLP tasks, adapting these models to specialized domains through conventional full fine-tuning remains prohibitively expensive. Modern foundation models typically contain billions of parameters, requiring substantial GPU memory, computational resources, and optimization overhead during training. For many academic researchers and industrial practitioners, fully updating all model parameters for every downstream financial task is neither economically practical nor computationally scalable. These limitations have shifted recent research from maximizing model size toward improving adaptation efficiency, giving rise to the paradigm of parameter-efficient fine-tuning (PEFT) \cite{han2024parameterefficientfinetuninglargemodels}. 

Rather than modifying the entire pretrained model, PEFT methods adapt only a small subset of trainable parameters while keeping the backbone model largely frozen \cite{han2024parameterefficientfinetuninglargemodels}. This strategy dramatically reduces memory consumption, training time, and storage requirements while preserving most of the capabilities acquired during large-scale pretraining. As a result, PEFT has become one of the dominant adaptation paradigms for large language models, particularly in domain-specific applications such as finance, where computational resources and high-quality labeled datasets are often limited. More importantly, PEFT enables researchers to efficiently customize foundation models for diverse downstream tasks without sacrificing scalability, making it a natural foundation for the next generation of financial language models. 

The development of parameter-efficient fine-tuning has itself followed a clear evolutionary trajectory, with successive methods addressing increasingly practical challenges in adapting large language models. LoRA (Low-Rank Adaptation) introduced the key insight that updating the full parameter matrix during fine-tuning is often unnecessary \cite{hu2021loralowrankadaptationlarge}. Instead, LoRA freezes the pretrained model weights and learns task-specific updates through a pair of trainable low-rank matrices, thereby reducing the number of trainable parameters by several orders of magnitude while maintaining competitive downstream performance. This simple yet effective formulation established LoRA as the foundation of modern PEFT methods and has since become one of the most widely adopted adaptation techniques for large language models.

Building upon LoRA, QLoRA addressed a different bottleneck—GPU memory consumption \cite{dettmers2023qloraefficientfinetuningquantized}. Rather than storing the backbone model in full precision, QLoRA performs backpropagation through a frozen 4-bit quantized model while training LoRA adapters in higher precision. Combined with innovations such as NormalFloat-4 (NF4) quantization, double quantization, and paged optimizers, QLoRA dramatically reduces memory requirements without sacrificing task performance, enabling models with tens of billions of parameters to be fine-tuned on a single consumer GPU \cite{dettmers2023qloraefficientfinetuningquantized}. This breakthrough substantially democratized large language model research and has become one of the most influential PEFT techniques for resource-constrained applications.

Subsequent research has focused on further improving the efficiency and effectiveness of parameter allocation and quantization. AdaLoRA observes that different layers contribute unequally during adaptation and therefore dynamically reallocates the low-rank budget according to parameter importance instead of assigning a uniform rank across all layers \cite{zhang2023adaloraadaptivebudgetallocation}. This adaptive strategy enables better utilization of a fixed parameter budget and often achieves superior performance under the same computational constraints. LoftQ further extends this line of research by recognizing that quantization and low-rank adaptation should be optimized jointly rather than independently \cite{li2023loftqlorafinetuningawarequantizationlarge}. Instead of directly applying LoRA to an already quantized model, LoftQ introduces a LoRA-aware quantization strategy that minimizes quantization error during initialization, thereby reducing the performance gap between quantized PEFT models and full-precision fine-tuning. Collectively, these methods illustrate a clear progression in PEFT research—from reducing trainable parameters, to minimizing memory consumption, to optimizing parameter allocation and quantization quality—providing increasingly practical solutions for adapting foundation models under limited computational resources. 

\subsection{Research Gap}
Over the past decade, financial sentiment analysis has evolved from lexicon-based sentiment estimation to deep contextual language understanding, accompanied by parallel advances in benchmark datasets and modeling approaches. Benchmark construction has expanded from sentence-level financial news classification to more diverse settings involving social media, monetary policy communications, and entity-level sentiment annotation. Meanwhile, methodological development has progressed from manually designed financial lexicons and traditional machine learning algorithms to transformer-based pretrained language models and, more recently, foundation-scale large language models. These advances have substantially improved the ability of NLP systems to capture contextual semantics and domain-specific financial knowledge while broadening the range of downstream financial applications.

The emergence of open-source financial LLMs and parameter-efficient adaptation techniques has further accelerated research by making large-scale language models accessible to a much wider community. Rather than focusing solely on developing increasingly larger models, recent studies have emphasized efficient adaptation, standardized evaluation, and practical deployment under realistic computational constraints. Collectively, these trends indicate that financial sentiment analysis has entered a new stage in which model adaptability, computational efficiency, and cross-domain generalization have become equally important research objectives.

Despite these significant advances, several research gaps remain. First, many existing studies evaluate models on a single benchmark dataset or a narrow range of financial text sources, making it difficult to assess whether performance improvements generalize across heterogeneous financial scenarios. Differences in annotation schemes, sentiment taxonomies, and domain characteristics further complicate direct comparisons between published results. Second, although recent work has introduced increasingly capable financial LLMs, comparative evaluations across the full methodological spectrum—from traditional machine learning and domain-specific pretrained language models to modern foundation models—remain relatively limited under unified experimental settings. Finally, while parameter-efficient fine-tuning has emerged as a practical solution for adapting large language models, most existing studies focus on demonstrating the effectiveness of individual adaptation techniques or financial LLM frameworks rather than systematically evaluating their applicability to financial sentiment classification across multiple benchmark datasets.

These observations suggest that evaluation should distinguish linguistic
performance from downstream financial validity. Motivated by this gap, the
present study first compares traditional machine learning, off-the-shelf
domain-specific encoders, zero-shot Qwen2.5, and QLoRA-adapted LLMs under a
unified three-class classification protocol. It then conducts a separate
return-based comparison among the seven probability-producing specifications,
excluding autoregressive zero-shot Qwen, on the same out-of-sample
news--stock panel.
This two-stage design tests both efficient financial-language adaptation and
the extent to which sentiment rankings survive a change in objective and data.

\subsection{Motivations}
The success of parameter-efficient adaptation has rapidly accelerated its adoption in financial NLP, where proprietary data, long financial documents, and limited computational resources make efficient model customization particularly valuable. Rather than developing entirely new financial foundation models, recent studies increasingly focus on adapting general-purpose LLMs to specialized financial tasks through lightweight fine-tuning strategies. This shift enables practitioners to leverage the extensive linguistic knowledge acquired during large-scale pretraining while efficiently incorporating domain-specific financial expertise, significantly reducing both computational cost and deployment barriers.

Representative studies, such as FinLoRA, demonstrate that parameter-efficient adaptation is not only computationally attractive but also highly effective across a broad range of financial applications. By systematically benchmarking multiple LoRA variants on diverse financial tasks, including sentiment analysis, question answering, SEC filing analysis, and financial information extraction, FinLoRA shows that PEFT methods consistently improve downstream performance while maintaining practical memory and training efficiency. More importantly, these studies \cite{wang2025finlorafinetuningquantizedfinancial, hu2021loralowrankadaptationlarge, dettmers2023qloraefficientfinetuningquantized, zhang2023adaloraadaptivebudgetallocation} suggest that the future of financial LLM research is increasingly centered on efficient adaptation and systematic evaluation rather than training increasingly larger domain-specific foundation models from scratch. 

Motivated by these developments, an important research question remains: to
what extent can parameter-efficient adaptation bridge the gap between
general-purpose foundation models and specialized financial sentiment
classification under realistic computational constraints? A related question
is whether improvements against human labels carry over to future-return
rankings. Addressing both requires a consistent supervised benchmark followed
by a separate downstream test, rather than treating linguistic and economic
performance as interchangeable measures.

\section{Methodology}

\subsection{Research Design}

The empirical study contains two experiments with different data and different
objectives. Experiment~1 is a supervised three-class sentiment benchmark. It
measures agreement with human or source-provided labels and is used to assess
the effect of QLoRA adaptation, backbone choice, and class-weighted training.
Experiment~2 is a temporally separate downstream evaluation. It applies frozen
classifiers to unlabeled financial news and tests whether their continuous
sentiment scores rank subsequent stock returns. Separating the two experiments
prevents downstream price movements from affecting model training and avoids
interpreting classification accuracy as direct evidence of tradable alpha.

\subsection{Experiment 1: Financial Sentiment Benchmark}

\subsubsection{Datasets and Label Harmonization}

The supervised benchmark combines five sources: Financial PhraseBank (FPB),
the FOMC monetary-policy corpus, SEntFiN~1.0, Twitter Financial News Sentiment,
and NASDAQ financial news \cite{malo2014GoodDO,shah-etal-2023-trillion,
Sinha_2022}. Together they cover conventional financial news, entity-specific
headlines, policy communication, and short social-media text. All observations
are represented as a text--label pair and harmonized to
$y\in\{-1,0,+1\}$ for negative, neutral, and positive sentiment.

Native negative/neutral/positive annotations in FPB and SEntFiN are retained.
For Twitter Financial News Sentiment, the released integer labels are mapped
from bearish, bullish, and neutral to $-1$, $+1$, and $0$, respectively. FOMC
policy stances are mapped from hawkish, neutral, and dovish to the benchmark's
negative, neutral, and positive directions. For the NASDAQ source, ratings of
0--1 are treated as negative, ratings of 2--3 as neutral, and ratings of 4--5
as positive. The mapping produces a common prediction space, but it does not
imply that the original annotation tasks are semantically identical; source
heterogeneity is intentionally retained as part of the robustness challenge.

FiQA is excluded even though it is widely used in financial NLP \cite{fiqa}.
Its continuous sentiment scores would require researcher-selected cutoffs to
create three discrete classes, making the resulting labels sensitive to an
additional thresholding decision. The final consolidated benchmark contains
33,549 observations.

\subsubsection{Train, Validation, and Test Splits}

The consolidated sample is partitioned by stratified random sampling so that
the three class proportions remain stable. A fixed seed of 42 is used. The
held-out test set contains 15\% of all observations; the remaining development
sample is divided into training and validation sets, yielding effective shares
of 76.5\%, 8.5\%, and 15.0\%. Validation data are used for model selection and
the test set is reserved for final reporting. Table~\ref{tab:data_split} gives
the resulting counts.

\begin{table}[t]
\centering
\caption{Experiment~1 partition statistics after stratified sampling.}
\label{tab:data_split}
\small
\begin{tabular}{lrrrr}
\toprule
Split & Samples & Negative & Neutral & Positive \\
\midrule
Training   & 25,664 & 20.37\% & 51.64\% & 27.99\% \\
Validation &  2,852 & 20.51\% & 50.98\% & 28.51\% \\
Test       &  5,033 & 20.39\% & 51.58\% & 28.03\% \\
\bottomrule
\end{tabular}
\end{table}

\subsection{Model Specifications}

The model rosters differ across the two experiments only because the
autoregressive zero-shot Qwen specification does not provide probability
vectors that are directly comparable with the sequence-classification models.
Experiment~1 therefore contains eight reported specifications: TF--IDF Naive
Bayes, FinBERT, Financial-RoBERTa, zero-shot Qwen2.5, and four QLoRA variants.
Experiment~2 contains the same probability-producing models but excludes
zero-shot Qwen, resulting in seven downstream specifications. FinBERT and
Financial-RoBERTa are evaluated in both experiments using their publicly
available sentiment-classification checkpoints without additional fine-tuning
on the consolidated Experiment~1 training split.
Table~\ref{tab:benchmark_models} summarizes this distinction.

\begin{table*}[t]
\centering
\caption{Model specifications used in the two experiments. A check mark means
that the model appears in the corresponding reported results.}
\label{tab:benchmark_models}
\small
\begin{tabular}{lllcc}
\toprule
Category & Model specification & Adaptation or inference & Exp.~1 & Exp.~2 \\
\midrule
Traditional ML & TF--IDF + Multinomial Naive Bayes & Supervised fit
& $\checkmark$ & $\checkmark$ \\
Domain PLM & FinBERT & Pretrained classification head
& -- & $\checkmark$ \\
Domain PLM & Financial-RoBERTa & Pretrained classification head
& -- & $\checkmark$ \\
Vanilla LLM & Qwen2.5-7B-Instruct & Zero-shot prompting
& $\checkmark$ & -- \\
PEFT LLM & Qwen2.5-7B-Instruct & QLoRA
& $\checkmark$ & $\checkmark$ \\
PEFT LLM & Qwen2.5-7B-Instruct & QLoRA + weighted CE
& $\checkmark$ & $\checkmark$ \\
PEFT LLM & LLaMA3-8B-Instruct & QLoRA
& $\checkmark$ & $\checkmark$ \\
PEFT LLM & Mistral-7B-Instruct & QLoRA
& $\checkmark$ & $\checkmark$ \\
Domain PLM & FinBERT & Off-the-shelf sentiment checkpoint
& $\checkmark$ & $\checkmark$ \\
Domain PLM & Financial-RoBERTa & Off-the-shelf sentiment checkpoint
& $\checkmark$ & $\checkmark$ \\
\bottomrule
\end{tabular}
\end{table*}

\subsubsection{Traditional and Domain-Specific Baselines}

The traditional supervised baseline applies TF--IDF text features followed by
a Multinomial Naive Bayes classifier fitted on the Experiment~1 training split.
It provides a computationally inexpensive reference in both experiments.

FinBERT and Financial-RoBERTa represent compact domain-specific Transformer
encoders \cite{araci2019finbertfinancialsentimentanalysis,shaikh2023query}.
They are evaluated using their publicly available three-class financial
sentiment checkpoints without additional fine-tuning on the consolidated
training data. Their Experiment~1 results should therefore be interpreted as
off-the-shelf transfer references rather than as a controlled architecture
comparison with the QLoRA models. The same frozen checkpoints are used to
generate continuous sentiment probabilities in Experiment~2.

\subsubsection{Vanilla and QLoRA-Adapted LLMs}

Qwen2.5-7B-Instruct is first evaluated by zero-shot prompting and constrained to
return one of the three sentiment classes. This supplies a same-backbone
reference for measuring the gain from task-specific adaptation. Qwen2.5-7B,
LLaMA3-8B, and Mistral-7B are then adapted as three-class sequence classifiers
with QLoRA. The pretrained backbone is stored in 4-bit form and frozen, while
low-rank adapters are inserted into the attention and feed-forward projection
layers. The common adapter settings are shown in
Table~\ref{tab:qlora_config}.

\begin{table}[t]
\centering
\caption{QLoRA configuration used for the three LLM backbones.}
\label{tab:qlora_config}
\begin{tabular}{ll}
\toprule
Hyperparameter & Value \\
\midrule
Task & Sequence classification \\
Quantization & 4-bit \\
LoRA rank ($r$) & 16 \\
LoRA scaling ($\alpha$) & 32 \\
LoRA dropout & 0.05 \\
Bias & None \\
Target modules & $q,k,v,o$ projections \\
 & gate, up, down projections \\
\bottomrule
\end{tabular}
\end{table}

All three backbones use the common optimization settings in
Table~\ref{tab:training_config}. A three-class classification head converts the
final representation into logits, followed by a softmax for class
probabilities. Models are selected using validation macro-F1 and evaluated once
on the held-out test split.

\begin{table}[t]
\centering
\caption{Optimization settings for QLoRA fine-tuning.}
\label{tab:training_config}
\begin{tabular}{ll}
\toprule
Hyperparameter & Value \\
\midrule
Optimizer & Paged AdamW (8-bit) \\
Learning rate & $2\times10^{-5}$ \\
Epochs & 3 \\
Maximum sequence length & 512 \\
Training batch size & 1 \\
Evaluation batch size & 2 \\
Gradient accumulation & 32 steps \\
Warmup ratio & 0.03 \\
Weight decay & 0.01 \\
Mixed precision & BF16 \\
Random seed & 42 \\
\bottomrule
\end{tabular}
\end{table}

For the Qwen2.5 loss ablation, the standard cross-entropy objective is compared
with inverse-frequency class-weighted cross-entropy:
\begin{equation}
\mathcal L_{\mathrm{WCE}}
=-\frac{1}{B}\sum_{b=1}^{B}w_{y_b}\log p_{b,y_b},
\qquad
w_c=\frac{N}{C N_c},
\label{eq:weighted_ce}
\end{equation}
where $B$ is the batch size, $C=3$, $N_c$ is the number of training examples in
class $c$, and $p_{b,y_b}$ is the probability assigned to the correct class.
This ablation changes the training objective, not the downstream probability
score or inference architecture.

The experiments use Hugging Face Transformers and PEFT with BitsAndBytes
quantization. Fine-tuning and recorded validation evaluations are performed on
NVIDIA L4 GPUs using BF16 arithmetic for trainable operations.

\subsection{Experiment 1 Evaluation Protocol}

Macro-F1 is the primary classification metric because the neutral class makes
up slightly more than half of the benchmark. For class $c$, precision, recall,
and F1 are computed in the usual one-versus-rest form, and macro-F1 is
\begin{equation}
\mathrm{MacroF1}=\frac{1}{3}\sum_{c\in\{-1,0,+1\}}
\frac{2\,\mathrm{Precision}_c\,\mathrm{Recall}_c}
{\mathrm{Precision}_c+\mathrm{Recall}_c}.
\end{equation}
We also report accuracy, macro-precision, macro-recall, weighted-F1, and
class-wise F1. Confusion matrices are inspected to identify whether errors
occur between directional classes or between a directional class and neutral;
the corresponding error counts are summarized in the results rather than
presented as a complete set of matrix figures.

\subsection{Experiment 2: Downstream Financial Evaluation}
\label{sec:downstream}

\begin{figure*}[t]
    \centering
    \resizebox{0.96\textwidth}{!}{
    \begin{tikzpicture}[
        node distance=0.7cm and 0.8cm,
        >=Latex,
        box/.style={rectangle,rounded corners=2pt,draw,align=center,
        minimum width=3.25cm,minimum height=1.15cm,font=\small},
        arrow/.style={->,thick}
    ]
    \node[box] (data) {\textbf{Benzinga 2019}\\10,637 unique headlines\\
        Fixed S\&P 100};
    \node[box, right=of data] (inference) {\textbf{Seven Classifiers}\\
        $p^{\mathrm{neg}},p^{\mathrm{neu}},p^{\mathrm{pos}}$};
    \node[box, right=of inference] (score) {\textbf{Continuous Score}\\
        $s=p^{\mathrm{pos}}-p^{\mathrm{neg}}$};
    \node[box, right=of score] (aggregate) {\textbf{Fresh-News Signal}\\
        Stock--date mean\\No-news stocks excluded};
    \node[box, below=1.0cm of aggregate] (alignment) {\textbf{Return Alignment}\\
        Next-session open\\$h=1,2,3,5$ days};
    \node[box, left=of alignment] (rank) {\textbf{Cross-Sectional Tests}\\
        Rank IC and ICIR\\NW + FDR inference};
    \node[box, left=of rank] (portfolio) {\textbf{Portfolio Tests}\\
        Equal-weight top/bottom 15\%\\Overlapping cohorts};
    \node[box, left=of portfolio] (metrics) {\textbf{Gross Performance}\\
        Return, Sharpe ratio,\\maximum drawdown};

    \draw[arrow] (data) -- (inference);
    \draw[arrow] (inference) -- (score);
    \draw[arrow] (score) -- (aggregate);
    \draw[arrow] (aggregate) -- (alignment);
    \draw[arrow] (alignment) -- (rank);
    \draw[arrow] (rank) -- (portfolio);
    \draw[arrow] (portfolio) -- (metrics);
    \end{tikzpicture}
    }
    \caption{Experiment~2 pipeline. The same fresh-news stock--date sample is
    used for model-specific Rank IC tests and equal-weighted portfolio tests.}
    \label{fig:downstream_pipeline}
\end{figure*}
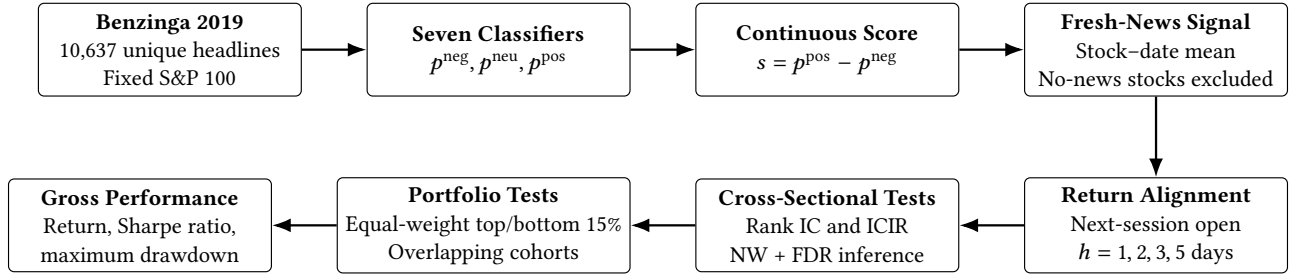

Experiment~2 evaluates economic validity rather than additional sentiment-label
accuracy. Figure~\ref{fig:downstream_pipeline} summarizes the final design.

\subsubsection{News Sample and Sentiment Inference}
\label{subsec:benzinga_downstream_data}

The downstream sample is drawn from the Benzinga analyst-ratings data and is
both temporally and substantively separate from Experiment~1. The universe is
fixed to the historical constituents of the S\&P~100 as of January~1, 2019,
and the sample covers January~1 through December~31, 2019. Headline or title
text is used; the separate partner-headlines file is excluded. Filtering to the
fixed universe and period yields 13,115 headline--stock observations. Because
one article can refer to several securities, these observations correspond to
10,637 unique URLs across 253 calendar dates.

A stable \texttt{news\_id} is assigned to each unique URL, and each model runs
inference once per unique headline. Predictions are then mapped back to the
13,115 headline--stock associations before stock-level aggregation. This avoids
repeating model inference for multi-stock articles while preserving every
security association in the economic test.

For model $m$ and headline $n$, the three output probabilities are converted
to a continuous expected-polarity score:
\begin{equation}
s_{n,m}=p^{\mathrm{pos}}_{n,m}-p^{\mathrm{neg}}_{n,m}\in[-1,1].
\label{eq:continuous_sentiment}
\end{equation}
Neutral has value zero. The continuous score retains both direction and model
confidence, but probabilities are not assumed to be calibrated across model
families. Accordingly, the analysis ranks signals separately within each model
instead of applying a common absolute threshold. The seven models are compared
on their common available sample.

\subsubsection{Stock--Date Aggregation and Return Alignment}

For stock $i$ on calendar date $d$, headline scores are averaged:
\begin{equation}
\bar{s}_{i,d,m}=\frac{1}{N_{i,d}}
\sum_{n\in\mathcal N_{i,d}}s_{n,m},
\label{eq:daily_sentiment}
\end{equation}
where $\mathcal N_{i,d}$ is the set of headlines associated with stock $i$ on
date $d$. Averaging prevents a frequently covered firm from receiving a larger
signal merely because it has more headlines. Only stock--date observations with
fresh news enter the daily cross-section; stocks without news are excluded and
are not assigned tied zero signals.

Reliable intraday publication times are unavailable for the full sample.
Therefore, every calendar-date signal is conservatively mapped to the first
trading session $t$ strictly after date $d$. Entry occurs at the adjusted open
of $t$. For $h\in\{1,2,3,5\}$ trading days, the aligned return is
\begin{equation}
R^{(h)}_{i,t}
=\frac{P^{\mathrm{close}}_{i,t+h-1}}
{P^{\mathrm{open}}_{i,t}}-1.
\label{eq:multi_horizon_return_method}
\end{equation}
Thus the one-day outcome is an open-to-close return, and longer horizons run
from the same entry open through the close of the $h$th session. This timing
rule avoids using same-date price movements that may precede the headline, but
it may also miss rapid intraday price discovery.

\subsubsection{Rank IC and Statistical Inference}

For every entry session with a valid cross-section, model-specific predictive
content is measured by the Spearman correlation between sentiment rank and
subsequent return:
\begin{equation}
\mathrm{IC}^{(h)}_{m,t}
=\operatorname{SpearmanCorr}_{i}
\left(\bar{s}_{i,t,m},R^{(h)}_{i,t}\right).
\label{eq:rank_ic}
\end{equation}
The reported Rank IC is the time-series mean of daily ICs. The annualized
information coefficient ratio is
\begin{equation}
\mathrm{ICIR}^{(h)}_m
=\frac{\overline{\mathrm{IC}}^{(h)}_m}
{\sigma(\mathrm{IC}^{(h)}_{m,t})}\sqrt{252}.
\label{eq:icir}
\end{equation}
Because adjacent multi-day outcomes overlap, tests of the mean IC use
Newey--West heteroskedasticity- and autocorrelation-consistent standard errors
with horizon-dependent lags. Benjamini--Hochberg false-discovery-rate (FDR)
correction is then applied jointly to the 28 model--horizon tests
(seven models by four horizons).

\subsubsection{Equal-Weighted Portfolio Tests}

Eligible stocks are ranked separately for each model and entry date. The most
positive 15\% form the long set $\mathcal L_{m,t}$ and the most negative 15\%
form the short set $\mathcal S_{m,t}$. The long-only and short-only legs use
unit gross exposure and equal weights within a leg. The long--short portfolio
allocates $+0.5$ to the long leg and $-0.5$ to the short leg, so that net
exposure is zero and total gross exposure is one:
\begin{equation}
w^{\mathrm{LS}}_{i,m,t}=
\begin{cases}
\displaystyle \frac{0.5}{|\mathcal L_{m,t}|},
&i\in\mathcal L_{m,t},\\[6pt]
\displaystyle -\frac{0.5}{|\mathcal S_{m,t}|},
&i\in\mathcal S_{m,t},\\[6pt]
0,&\text{otherwise}.
\end{cases}
\label{eq:long_short_weights}
\end{equation}

A new cohort is formed each entry session. One-day cohorts are held from the
entry open to that session's close. For $h>1$, each cohort is held for $h$
sessions and receives $1/h$ of portfolio capital, so up to $h$ cohorts overlap.
This construction converts overlapping holding periods into a daily portfolio
return series without treating multi-day observations as independent one-day
trades. We report gross total return, annualized Sharpe ratio, and maximum
drawdown. Commissions, bid--ask spreads, market impact, slippage, and stock-
borrow fees are excluded, so the portfolios are signal diagnostics rather than
estimates of net implementable performance.

\section{Experiment 1: Financial Sentiment Classification Results}

\begin{table*}[t]
\centering
\caption{Overall performance on the unified Experiment~1 test set
(\(N=5{,}033\)). FinBERT and Financial-RoBERTa use publicly available
sentiment checkpoints without additional adaptation to the consolidated
training split and are reported as off-the-shelf references.}
\label{tab:overall_results}
\begin{tabular}{llccccc}
\toprule
\textbf{Category} &
\textbf{Model} &
\textbf{Accuracy} &
\textbf{Macro-P} &
\textbf{Macro-R} &
\textbf{Macro-F1} &
\textbf{Weighted-F1} \\
\midrule

Traditional ML
& TF--IDF + Naïve Bayes
& 0.6976
& 0.7393
& 0.6061
& 0.6334
& 0.6764 \\

Vanilla LLM
& Qwen2.5-7B-Instruct (Zero-shot)
& 0.7280
& 0.7140
& 0.7579
& 0.7274
& 0.7283 \\

Domain PLM
& FinBERT
& 0.6930
& 0.6715
& 0.6828
& 0.6753
& 0.6933 \\

Domain PLM
& Financial-RoBERTa
& 0.6622
& 0.6668
& 0.7152
& 0.6679
& 0.6592 \\

PEFT LLM
& Qwen2.5-7B + QLoRA
& 0.8683
& 0.8635
& 0.8595
& 0.8615
& 0.8682 \\

PEFT LLM
& Qwen2.5-7B + QLoRA (Weighted CE)
& 0.8667
& 0.8618
& 0.8575
& 0.8595
& 0.8666 \\

PEFT LLM
& LLaMA3-8B + QLoRA
& 0.8814
& 0.8755
& \textbf{0.8751}
& 0.8753
& 0.8814 \\

PEFT LLM
& Mistral-7B + QLoRA
& \textbf{0.8840}
& \textbf{0.8792}
& \textbf{0.8751}
& \textbf{0.8771}
& \textbf{0.8839} \\

\bottomrule
\end{tabular}
\end{table*}

\subsection{Overall Benchmark Performance}

Table~\ref{tab:overall_results} reports the overall test-set performance of the
evaluated financial sentiment classification models. Macro-F1 is treated as
the primary evaluation metric because the benchmark exhibits a moderately
imbalanced class distribution, with neutral samples accounting for more than
half of the test set.

The traditional TF--IDF and Naïve Bayes baseline achieved an accuracy of
0.6976 and a macro-F1 score of 0.6334. Although its macro-precision was
relatively high, its substantially lower macro-recall indicates that the
model failed to identify a large proportion of directional sentiment
instances, particularly within the negative and positive classes.

The two off-the-shelf domain-specific encoders produced moderate transfer
performance on the heterogeneous benchmark. FinBERT achieved an accuracy of
0.6930 and a macro-F1 score of 0.6753, outperforming Financial-RoBERTa, which
obtained an accuracy of 0.6622 and a macro-F1 score of 0.6679.
Financial-RoBERTa nevertheless achieved the higher macro-recall of 0.7152,
indicating stronger average class coverage but less balanced precision and
F1 performance.

All four QLoRA specifications considerably outperformed both the traditional
baseline and the off-the-shelf domain-specific encoders. Among the evaluated backbones, Mistral-7B achieved the
highest overall performance, with an accuracy of 0.8840 and a macro-F1 score
of 0.8771. LLaMA3-8B produced closely comparable results, reaching an accuracy
of 0.8814 and a macro-F1 score of 0.8753. The difference between these two
models was small, suggesting that both backbones provide similarly strong
financial sentiment representations under the unified QLoRA configuration.

The standard Qwen2.5 QLoRA model obtained a macro-F1 score of 0.8615, whereas
the class-weighted cross-entropy variant achieved 0.8595. Therefore, weighting
the training loss according to class frequency did not improve overall
classification performance in this setting. 

These differences should not be interpreted as evidence that decoder-based
LLMs are intrinsically superior to financial encoders. The QLoRA models were
trained directly on the consolidated training split, whereas FinBERT and
Financial-RoBERTa were evaluated without additional adaptation to that split.
The comparison therefore shows the value of supervised task-specific
adaptation relative to off-the-shelf transfer, rather than providing a fully
controlled comparison of model architectures.

\subsection{Class-wise Performance and Error Analysis}
\label{sec:classwise_results}

Table~\ref{tab:classwise_f1} reports the class-wise F1 scores of the evaluated
models. The traditional TF--IDF and Naïve Bayes baseline exhibited a pronounced
bias toward the majority neutral class. Although it achieved a neutral-class
F1 score of 0.7787, its F1 scores for negative and positive sentiment were only
0.5177 and 0.6037, respectively. Its confusion matrix further shows that 509
negative samples and 642 positive samples were incorrectly classified as
neutral, indicating that the model frequently failed to identify directional
sentiment expressions.

The domain-specific encoders displayed different error profiles. FinBERT
achieved the stronger neutral-class F1 score of 0.7449 but lower positive-class
performance of 0.6260. Financial-RoBERTa achieved a higher negative-class F1
score of 0.7033, while its neutral-class F1 decreased to 0.6423. Neither model
matched the class-wise balance of the QLoRA-adapted classifiers.

\begin{table}[t]
\centering
\caption{Class-wise F1 scores on the financial sentiment test set.}
\label{tab:classwise_f1}
\begin{tabular}{lccc}
\toprule
\textbf{Model} &
\textbf{Negative} &
\textbf{Neutral} &
\textbf{Positive} \\
\midrule
TF--IDF + Naïve Bayes
& 0.5177 & 0.7787 & 0.6037 \\

Qwen2.5-7B-Instruct (Zero-shot)
& 0.7312 & 0.7326 & 0.7182 \\

FinBERT
& 0.6550 & 0.7449 & 0.6260 \\

Financial-RoBERTa
& 0.7033 & 0.6423 & 0.6581 \\

Qwen2.5 + QLoRA
& 0.8523 & 0.8870 & 0.8451 \\

Qwen2.5 + QLoRA (Weighted CE)
& 0.8475 & 0.8856 & 0.8454 \\

LLaMA3-8B + QLoRA
& \textbf{0.8648} & 0.8978 & 0.8632 \\

Mistral-7B + QLoRA
& 0.8619 & \textbf{0.9009} & \textbf{0.8685} \\
\bottomrule
\end{tabular}
\end{table}

In contrast, all QLoRA-adapted LLMs achieved substantially more balanced
performance across the three sentiment categories, with class-wise F1 scores
above 0.84. LLaMA3-8B obtained the highest negative-class F1 score of 0.8648,
while Mistral-7B achieved the strongest performance for both neutral and
positive sentiment, with F1 scores of 0.9009 and 0.8685, respectively. These
results suggest that the strongest LLM backbones differ primarily in how their
remaining classification errors are distributed rather than exhibiting a
systematic failure on any individual sentiment category.

Applying class-weighted cross-entropy to Qwen2.5 did not improve minority-class
performance. Relative to the standard QLoRA model, the weighted-loss variant
reduced the negative-class F1 score from 0.8523 to 0.8475 and the neutral-class
F1 score from 0.8870 to 0.8856, while producing only a negligible improvement
in positive-class F1 from 0.8451 to 0.8454. This indicates that the moderate
class imbalance in the benchmark was not sufficiently severe to benefit from
the adopted inverse-frequency weighting scheme.

The confusion matrices reveal that most remaining errors produced by the LLM
models occurred between directional sentiment and the neutral category rather
than between negative and positive sentiment directly. For example, the
LLaMA3 and Mistral models each produced only 67 direct negative--positive
reversals across the 5,033 test samples, compared with 164 such reversals for
Naïve Bayes. This pattern suggests that contextual language models generally
capture sentiment polarity correctly, while ambiguous, weakly expressed, or
implicit financial sentiment remains more difficult to distinguish from
neutral language.

\subsection{Effectiveness of QLoRA Adaptation}
\label{sec:qlora_effectiveness}

To evaluate the effectiveness of parameter-efficient adaptation, we analyze
the results from three complementary perspectives. First, Qwen2.5-7B is
evaluated both under zero-shot prompting and after QLoRA fine-tuning, providing
a direct comparison using the same underlying backbone. Second, the three
QLoRA-adapted backbones are compared under an identical training configuration
to examine whether the effectiveness of parameter-efficient fine-tuning
generalizes across model families. Finally, we investigate whether
class-weighted cross-entropy provides additional benefits under the moderate
class imbalance present in the benchmark.

The direct comparison between zero-shot and QLoRA-adapted Qwen2.5 provides
clear evidence of the effectiveness of task-specific parameter-efficient
adaptation. The zero-shot model achieved an accuracy of 0.7280 and a macro-F1
score of 0.7274, whereas QLoRA fine-tuning increased these metrics to 0.8683
and 0.8615, respectively. This corresponds to absolute improvements of 14.03
percentage points in accuracy and 13.41 percentage points in macro-F1.

The zero-shot model exhibited a systematic tendency to over-predict
directional sentiment while under-predicting the neutral class. Although
neutral samples represented 51.58\% of the test set, only 41.45\% of the
model's predictions were neutral. In particular, 410 neutral samples were
classified as negative and 471 were classified as positive. QLoRA adaptation
substantially reduced these errors and produced a class distribution more
closely aligned with the underlying test data. The resulting improvement was
observed across all three sentiment categories, demonstrating that
task-specific adaptation enhanced both overall accuracy and class-wise
balance.

Under the unified QLoRA configuration, the three evaluated backbones achieved
consistently strong performance, although meaningful differences remained
across model families. Mistral-7B achieved the highest macro-F1 score of
0.8771, followed closely by LLaMA3-8B with 0.8753, while Qwen2.5-7B obtained
0.8615. The relatively small difference between Mistral and LLaMA3 suggests
that both architectures adapt effectively to financial sentiment
classification under the same parameter-efficient training protocol.
Nevertheless, the lower performance of Qwen2.5 indicates that the benefits of
QLoRA remain backbone-dependent rather than entirely architecture-independent.

The comparison between the two Qwen2.5 variants shows that class-weighted
cross-entropy did not improve performance. The standard-loss model achieved a
macro-F1 score of 0.8615, whereas the weighted-loss variant obtained 0.8595.
Similarly, accuracy decreased slightly from 0.8683 to 0.8667. Class-wise
analysis shows that weighting produced no meaningful improvement for the
positive class and slightly reduced performance for the negative and neutral
classes. These findings suggest that the benchmark's class imbalance was not
severe enough to justify the adopted inverse-frequency weighting scheme.
\begin{table}[t]
\centering
\caption{Effect of QLoRA adaptation and loss weighting on Qwen2.5.}
\label{tab:qwen_ablation}
\begin{tabular}{lccc}
\toprule
\textbf{Qwen2.5 Variant} &
\textbf{Accuracy} &
\textbf{Macro-F1} &
\textbf{Weighted-F1} \\
\midrule

Zero-shot prompting
& 0.7280
& 0.7274
& 0.7283 \\

QLoRA
& \textbf{0.8683}
& \textbf{0.8615}
& \textbf{0.8682} \\

QLoRA + Weighted CE
& 0.8667
& 0.8595
& 0.8666 \\

\bottomrule
\end{tabular}
\end{table}

\subsection{Computational Efficiency}
\label{sec:computational_efficiency}

Beyond predictive performance, we report the observed evaluation runtime and
throughput of the QLoRA-adapted models on the validation set. Each recorded run
used the same validation split containing 2,852 samples, and all QLoRA models
performed three-class sequence classification through a single forward pass.
However, the measurements were collected from individual evaluation runs rather
than repeated under a fully controlled benchmarking protocol. Consequently,
the reported values should be interpreted as descriptive runtime observations
rather than precise hardware-level comparisons.

\begin{table}[t]
\centering
\caption{Observed validation-set evaluation runtime of the QLoRA-adapted models.}
\label{tab:computational_efficiency}
\small
\setlength{\tabcolsep}{2pt}
\begin{tabular}{@{}p{0.39\columnwidth}rrr@{}}
\toprule
\textbf{Model} &
\shortstack{\textbf{Runtime}\\\textbf{(s)}} &
\shortstack{\textbf{Throughput}\\\textbf{(samples/s)}} &
\shortstack{\textbf{Macro-}\\\textbf{F1}} \\ \\
\midrule

Qwen2.5-7B + QLoRA
& 226.25
& 12.606
& 0.8646 \\

Qwen2.5-7B + QLoRA (Weighted CE)
& 232.11
& 12.287
& 0.8652 \\

LLaMA3-8B + QLoRA
& 264.41
& 10.786
& 0.8767$^{\dagger}$ \\

Mistral-7B + QLoRA
& 262.19
& 10.878
& 0.8761 \\

\bottomrule
\multicolumn{4}{p{0.94\columnwidth}}{\footnotesize
$^{\dagger}$The runtime and macro-F1 were recorded during the original
Trainer-based validation run. A later manual re-evaluation obtained a
macro-F1 score of 0.8756 but did not record evaluation runtime.}
\end{tabular}
\end{table}

As shown in Table~\ref{tab:computational_efficiency}, the two Qwen2.5 variants
recorded the shortest evaluation runtimes, completing the validation set in
approximately 226--232 seconds. Their observed throughput rates were also
similar, at approximately 12.3--12.6 samples per second. This similarity is
expected because class-weighted cross-entropy modifies the training objective
but does not alter the model architecture or the inference procedure. The
small runtime difference between the two runs is therefore more plausibly
attributable to normal system-level variation than to the loss function itself.

The recorded evaluation times for LLaMA3-8B and Mistral-7B were approximately
264 and 262 seconds, corresponding to throughput rates of approximately
10.8 samples per second. These models also achieved higher validation
macro-F1 scores than the Qwen2.5 variants in the recorded runs. The observations
therefore suggest a possible efficiency--performance trade-off, in which
Qwen2.5 provides faster evaluation while LLaMA3 and Mistral provide stronger
classification performance. Nevertheless, because the measurements were not
repeated across multiple runs and transient hardware or system conditions were
not fully controlled, these runtime differences should not be interpreted as
statistically significant.

The zero-shot Qwen2.5 baseline is excluded from the runtime comparison because
it generates sentiment labels autoregressively, whereas the QLoRA models
produce class logits through a sequence-classification head. These two
inference procedures have fundamentally different computational characteristics
and are therefore not directly comparable. In addition, part of the zero-shot
evaluation was affected by a temporary GPU-driver failure, further preventing
a controlled runtime comparison.

Peak GPU memory consumption, total fine-tuning time, trainable parameter
counts, and adapter storage requirements were not logged consistently across
all experiments. Accordingly, this section provides a descriptive comparison
of observed validation runtime and throughput rather than a comprehensive
assessment of training or deployment efficiency.

\section{Experiment 2: Downstream Financial Evaluation}
\label{sec:downstream_results}

\subsection{Descriptive Comparison of Out-of-Sample Predictions}
\label{subsec:benzinga_prediction_results}

The following figures compare the prediction behavior of the seven models on
the 10,637 unique Benzinga news items. Because this downstream sample is
unlabeled, these results are descriptive and do not constitute an additional
classification-accuracy evaluation.

\begin{figure*}[t]
    \centering
    \includegraphics[width=0.88\textwidth]
    {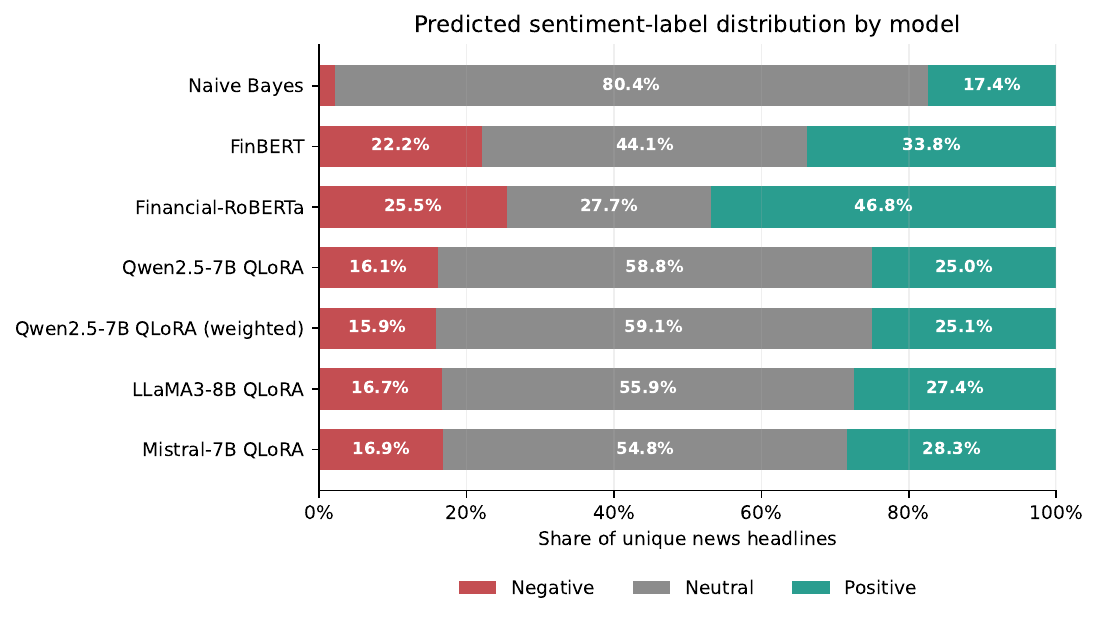}
    \caption{Predicted sentiment-label shares across seven models on the
    unlabeled 2019 Benzinga S\&P~100 unique-news sample
    ($N=10{,}637$). Each bar reports the proportion of headlines classified
    as negative, neutral, or positive.}
    \label{fig:benzinga_label_distribution}
    \Description{A stacked horizontal bar chart comparing the negative,
    neutral, and positive prediction shares of seven sentiment models.}
\end{figure*}

\begin{figure*}[t]
    \centering
    \includegraphics[width=0.88\textwidth]
    {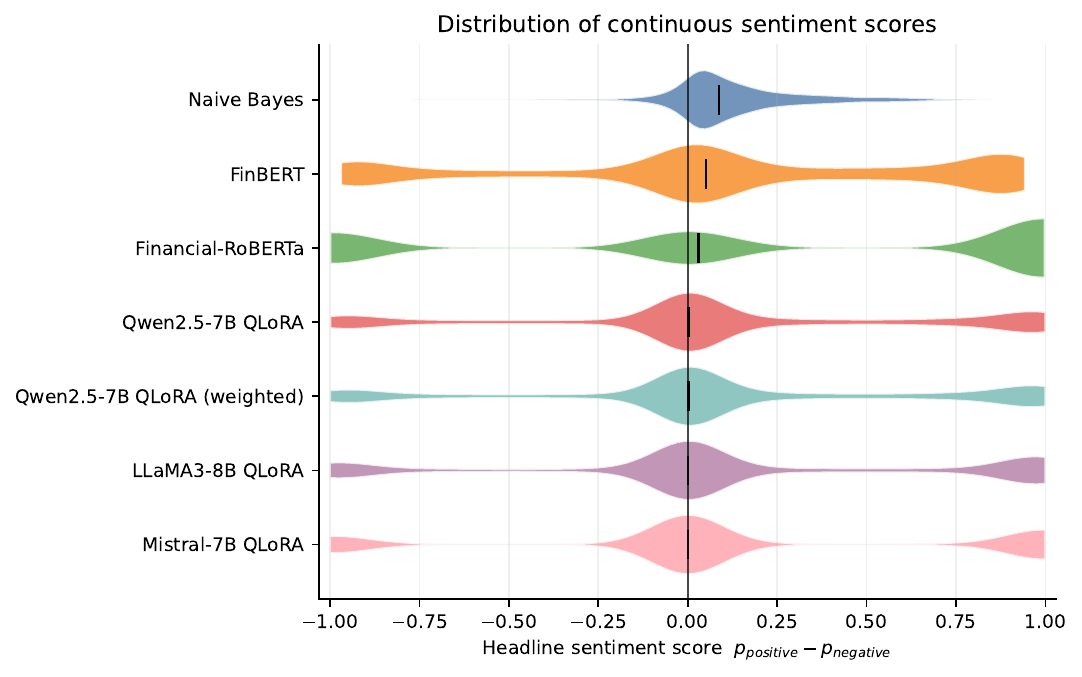}
    \caption{Model-specific distributions of continuous headline sentiment
    scores on the Benzinga sample. The score is defined as
    $s=p_{\mathrm{positive}}-p_{\mathrm{negative}}$, with values closer to
    $-1$ and $1$ indicating stronger negative and positive sentiment,
    respectively.}
    \label{fig:benzinga_score_distribution}
    \Description{Ridgeline distributions of continuous sentiment scores
    produced by seven sentiment models, ranging from negative one to
    positive one.}
\end{figure*}

\begin{figure*}[t]
    \centering
    \includegraphics[width=0.88\textwidth]
    {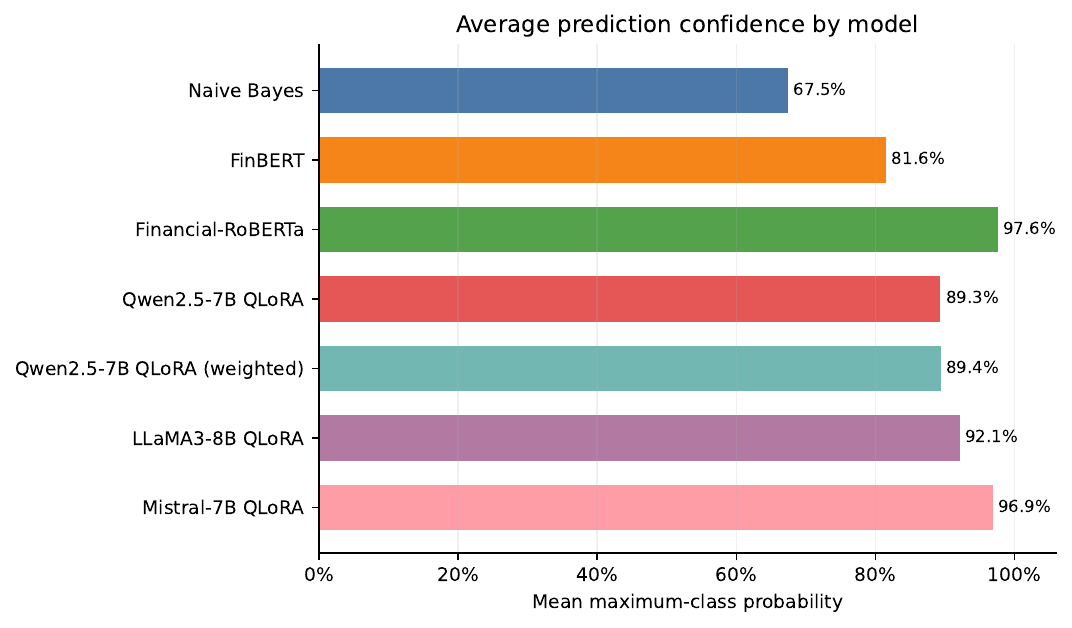}
    \caption{Mean prediction confidence of each model on the Benzinga sample,
    measured as the maximum predicted probability across the negative,
    neutral, and positive classes for each headline.}
    \label{fig:benzinga_mean_confidence}
    \Description{A horizontal bar chart comparing the mean maximum-class
    probability of seven sentiment models.}
\end{figure*}

Figures~\ref{fig:benzinga_label_distribution}--\ref{fig:benzinga_mean_confidence}
reveal substantial heterogeneity in model behavior. Naive Bayes assigns
80.4\% of headlines to the neutral class, produces the most concentrated
continuous-score distribution, and has the lowest mean confidence at 67.5\%.
In contrast, Financial-RoBERTa assigns 46.8\% of headlines to the positive
class, places substantially more probability mass near the extremes of the
sentiment-score range, and records the highest mean confidence at 97.6\%.
The QLoRA-based models occupy an intermediate range, while the two Qwen2.5
variants are nearly indistinguishable in their label shares, score
distributions, and confidence levels; their continuous scores have a Pearson
correlation of 0.998 and their discrete predictions agree on 98.7\% of
headlines. However, higher confidence should not be interpreted as higher
accuracy because the Benzinga sample has no gold sentiment labels and the
predicted probabilities are not calibrated on it. These model-specific
differences motivate the use of within-model daily cross-sectional rankings
in the subsequent trading analysis rather than a common absolute sentiment
threshold.

\subsection{Return-Based Economic Evaluation}
\label{subsec:downstream_financial_evaluation}

The supervised classification results measure whether a model reproduces human
sentiment labels, but do not establish that its outputs contain economically
useful information. We therefore conduct a separate downstream evaluation on
the 2019 Benzinga sample described in
Section~\ref{subsec:benzinga_downstream_data}. The experiment compares all seven
classifiers on a common stock--date sample and asks two related questions:
(i) whether the continuous sentiment scores rank subsequent stock returns, and
(ii) whether the extreme positive and negative signals support systematic
cross-sectional portfolios. This evaluation is deliberately treated as an
out-of-sample economic-validity test rather than an additional sentiment-label
benchmark.

The score construction, fresh-news sample restriction, next-session return
alignment, Rank IC inference, and overlapping-cohort portfolios follow the
protocol in Section~\ref{sec:downstream}. The results below therefore focus on
predictability and portfolio outcomes rather than repeating the experimental
definitions.

\subsubsection{Cross-sectional predictability}

\begin{figure*}[t]
    \centering
    \includegraphics[width=\textwidth]{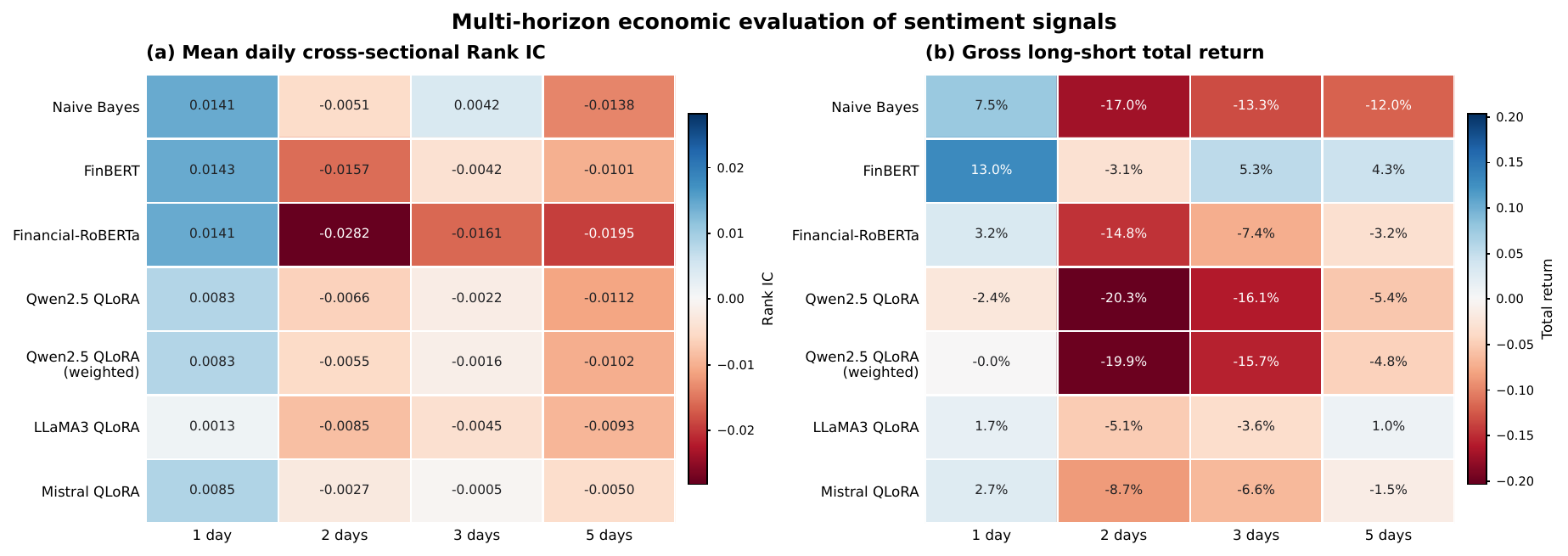}
    \caption{Multi-horizon downstream evaluation. Panel (a) reports the mean
    daily cross-sectional Spearman rank IC between fresh-news sentiment scores
    and subsequent returns. Panel (b) reports gross total returns of the
    equal-weighted long--short portfolios. Blue denotes positive and red denotes
    negative values. Statistical inference uses Newey--West standard errors and
    FDR correction; none of the 28 IC tests is significant after correction.}
    \label{fig:downstream_multi_horizon}
\end{figure*}

Figure~\ref{fig:downstream_multi_horizon}a and
Table~\ref{tab:multi_horizon_summary} show a common but weak horizon pattern.
At the one-day horizon, all seven models produce positive mean rank ICs, ranging
from 0.0013 for LLaMA3 QLoRA to 0.0143 for FinBERT. Financial-RoBERTa and Naive
Bayes are close behind at 0.0141, whereas Qwen2.5 QLoRA and its class-weighted
variant both reach 0.0083. The positive relationship does not persist: every
model has a negative two-day IC, most remain near zero or negative at three
days, and all are negative at five days. This sign pattern is consistent with
a small amount of immediate price continuation followed by rapid signal decay
or reversal.

The estimates are descriptive rather than statistically conclusive. The
largest one-day Newey--West annualized ICIR is 1.015 for Financial-RoBERTa,
followed by 0.972 for FinBERT and 0.951 for Naive Bayes. However, none of the 28
model--horizon tests remains significant after FDR adjustment: the minimum
adjusted $q$-value is 0.9622. Even Financial-RoBERTa's two-day estimate, whose
unadjusted Newey--West $p$-value is 0.0397, does not survive the correction.
Accordingly, the results do not establish robust return predictability for any
model.

\begin{table*}[t]
\centering
\caption{Mean Rank IC and gross long--short total return by model and holding
horizon. Returns are generated by overlapping cohorts for multi-day horizons.}
\label{tab:multi_horizon_summary}
\small
\setlength{\tabcolsep}{4.5pt}
\begin{tabular}{lrrrrrrrr}
\toprule
& \multicolumn{2}{c}{1 day} & \multicolumn{2}{c}{2 days}
& \multicolumn{2}{c}{3 days} & \multicolumn{2}{c}{5 days} \\
\cmidrule(lr){2-3}\cmidrule(lr){4-5}\cmidrule(lr){6-7}\cmidrule(lr){8-9}
Model & IC & L--S ret. & IC & L--S ret. & IC & L--S ret. & IC & L--S ret. \\
\midrule
Naive Bayes                 & 0.0141 &  7.47\% & -0.0051 & -17.01\% &  0.0042 & -13.31\% & -0.0138 & -11.96\% \\
FinBERT                     & 0.0143 & 12.96\% & -0.0157 &  -3.07\% & -0.0042 &   5.32\% & -0.0101 &   4.34\% \\
Financial-RoBERTa           & 0.0141 &  3.17\% & -0.0282 & -14.76\% & -0.0161 &  -7.36\% & -0.0195 &  -3.20\% \\
Qwen2.5 QLoRA              & 0.0083 & -2.35\% & -0.0066 & -20.34\% & -0.0022 & -16.07\% & -0.0112 &  -5.43\% \\
Qwen2.5 QLoRA (weighted)   & 0.0083 & -0.03\% & -0.0055 & -19.87\% & -0.0016 & -15.70\% & -0.0102 &  -4.76\% \\
LLaMA3 QLoRA               & 0.0013 &  1.69\% & -0.0085 &  -5.09\% & -0.0045 &  -3.61\% & -0.0093 &   0.98\% \\
Mistral QLoRA              & 0.0085 &  2.69\% & -0.0027 &  -8.66\% & -0.0005 &  -6.57\% & -0.0050 &  -1.51\% \\
\bottomrule
\end{tabular}
\end{table*}

\subsubsection{Portfolio performance and return decomposition}

The one-day horizon contains the clearest economically positive results. As
shown in Table~\ref{tab:one_day_portfolios}, FinBERT produces the strongest
market-neutral portfolio, with a gross total return of 12.96\%, a Sharpe ratio
of 1.11, and a maximum drawdown of 6.37\%. Its performance is driven primarily
by the short leg, which returns 19.73\% despite a cohort win rate below 50\%.
This combination indicates that the result reflects the magnitude of a smaller
number of successful short positions rather than uniformly accurate daily
directional forecasts. Naive Bayes ranks second on one-day long--short return
(7.47\%), while Financial-RoBERTa, Mistral QLoRA, and LLaMA3 QLoRA produce
smaller positive returns.

The two Qwen variants do not show a downstream advantage. The standard Qwen
model generates a one-day long--short return of $-2.35\%$, whereas class
weighting improves the result to approximately zero ($-0.03\%$) and reduces the
losses at the longer horizons. This is a modest relative improvement, but not
evidence of economically or statistically superior forecasting.

\begin{table*}[t]
\centering
\caption{One-day gross portfolio performance. Total return and Sharpe ratio are
reported for each portfolio leg; maximum drawdown is shown for the long--short
portfolio.}
\label{tab:one_day_portfolios}
\small
\setlength{\tabcolsep}{5pt}
\begin{tabular}{lrrrrrrr}
\toprule
& \multicolumn{2}{c}{Long-only} & \multicolumn{2}{c}{Short-only}
& \multicolumn{3}{c}{Long--short} \\
\cmidrule(lr){2-3}\cmidrule(lr){4-5}\cmidrule(lr){6-8}
Model & Return & Sharpe & Return & Sharpe & Return & Sharpe & Max DD \\
\midrule
Naive Bayes                 &  13.38\% &  0.73 &  -0.83\% &  0.13 &  7.47\% &  0.54 & -12.13\% \\
FinBERT                     &   4.05\% &  0.31 &  19.73\% &  0.96 & 12.96\% &  1.11 &  -6.37\% \\
Financial-RoBERTa           &  15.32\% &  1.01 &  -9.05\% & -0.37 &  3.17\% &  0.33 & -12.52\% \\
Qwen2.5 QLoRA              & -11.24\% & -0.59 &   5.10\% &  0.34 & -2.35\% & -0.14 & -12.42\% \\
Qwen2.5 QLoRA (weighted)   &  -8.92\% & -0.44 &   7.38\% &  0.45 & -0.03\% &  0.05 & -10.94\% \\
LLaMA3 QLoRA               &   6.52\% &  0.45 &  -5.05\% & -0.15 &  1.69\% &  0.20 & -11.46\% \\
Mistral QLoRA              &   2.74\% &  0.23 &   0.25\% &  0.11 &  2.69\% &  0.27 & -12.79\% \\
\bottomrule
\end{tabular}
\end{table*}

At longer horizons, the directional portfolios and the market-neutral
portfolios diverge. The long-only portfolios are broadly profitable at three
and five days---for example, FinBERT returns 39.64\% and 33.89\%, respectively,
with Sharpe ratios of 2.29 and 2.25---while the corresponding short-only
portfolios lose 22.15\% and 20.15\%. Most long--short portfolios are therefore
negative beyond one day. Because the ICs over the same horizons are near zero
or negative, this pattern is more consistent with positive market exposure in
the 2019 sample than with persistent cross-sectional sentiment alpha. FinBERT
is the main descriptive exception, retaining positive long--short returns of
5.32\% at three days and 4.34\% at five days, although its IC estimates remain
negative and statistically insignificant.

Taken together, the downstream results reveal a disconnect between sentiment
classification performance and economic predictability. Fine-tuned QLoRA
models can achieve strong classification accuracy without dominating the
return-based evaluation, while the strongest one-day trading result belongs to
FinBERT. Across model families, any economically useful sentiment effect is
small, concentrated in the first trading session, and not statistically robust
after multiple-testing correction. We therefore interpret Experiment~2 as an
economic-validity and signal-decay analysis, rather than evidence that one
sentiment classifier delivers reliable tradable alpha. Sample coverage,
exposure neutralization, trading frictions, and the absence of causal
identification are discussed in the concluding limitations.

\section{Conclusion and Limitations}
\label{sec:conclusion}

This study examined whether PEFT improves financial sentiment classification
and whether stronger linguistic performance translates into economically
meaningful return predictability. Experiment~1 compared QLoRA-adapted
Qwen2.5-7B, LLaMA3-8B, and Mistral-7B with zero-shot Qwen2.5, a TF--IDF
Naive Bayes baseline, and the off-the-shelf FinBERT and Financial-RoBERTa
encoders on a harmonized collection of five financial sentiment datasets.
Experiment~2 used a separate Benzinga sample to compare seven
probability-producing classifiers over multiple return horizons.

The classification results demonstrate that QLoRA provides an effective
parameter-efficient approach to financial adaptation. Fine-tuned Qwen2.5
achieved 86.83\% accuracy and 86.15\% macro-F1, improving macro-F1 by 13.41
percentage points over its zero-shot specification. Mistral-7B produced the
best overall test result, with 88.40\% accuracy and 87.71\% macro-F1, followed
closely by LLaMA3-8B. Inverse-frequency class weighting did not improve Qwen:
macro-F1 decreased from 86.15\% to 85.95\%. Thus, the main classification gain
comes from task-specific QLoRA adaptation rather than loss reweighting.

The off-the-shelf FinBERT and Financial-RoBERTa checkpoints achieved
macro-F1 scores of 67.53\% and 66.79\%, respectively, below all four
QLoRA specifications. However, this difference is not a controlled
architecture comparison because the QLoRA models were adapted to the
consolidated training split, whereas the encoder checkpoints were not.
The result therefore supports the effectiveness of task-specific QLoRA
adaptation relative to off-the-shelf transfer, but does not establish universal
superiority over domain-specific encoders.

Notably, the Experiment~1 classification ranking did not transfer directly to
Experiment~2. Despite its lower classification performance on the consolidated
benchmark, FinBERT generated the strongest one-day Rank IC and long--short
gross return. This contrast reinforces the distinction between reproducing
human sentiment labels and extracting economically predictive information.

The downstream financial evaluation yielded a more cautious result. The experiment used 10,637 unique Benzinga headlines corresponding to 13,115 stock--headline observations and evaluated sentiment signals over one-, two-, three-, and five-day return horizons. Across the seven model specifications, one-day Rank ICs were generally positive but economically small. FinBERT produced the largest one-day Rank IC of 0.0143, followed closely by Financial-RoBERTa and Naive Bayes, whereas the Qwen2.5 variants produced an IC of approximately 0.0083. None of the 28 model--horizon tests remained statistically significant after Newey--West adjustment and false-discovery-rate correction.

The horizon analysis further showed that the weak positive relationship was concentrated primarily within the first trading session. Rank ICs generally weakened or became negative at the two-, three-, and five-day horizons. This pattern is descriptively consistent with rapid information incorporation followed by signal decay or partial reversal. However, given the statistical uncertainty and limited evaluation sample, it should not be interpreted as conclusive evidence of either short-term continuation or subsequent return reversal.

Portfolio results were broadly consistent with this interpretation. FinBERT
generated the strongest one-day long--short gross return of 12.96\%, while the
standard Qwen2.5 QLoRA specification returned $-2.35\%$. The class-weighted
Qwen variant returned $-0.03\%$, a modest relative improvement that should not
be attributed to a different downstream scoring rule: both Qwen variants use
the same probability-weighted continuous sentiment score and differ only in
their training loss. At longer horizons, long-only portfolios frequently
generated positive returns while short-only portfolios generally lost money.
This asymmetry suggests that much of the apparent multi-day profitability was
associated with positive market exposure rather than persistent cross-sectional
sentiment alpha. FinBERT's stronger one-day portfolio was also unsupported by a
statistically significant overall IC and may have been influenced by a limited
number of large observations.

Taken together, the results reveal a distinction between sentiment-classification performance and downstream economic predictability. A model may accurately reproduce human sentiment labels without producing a sufficiently informative ranking of future returns. Sentiment classification and return prediction are related but fundamentally different objectives: the former measures semantic agreement with annotated labels, whereas the latter requires incremental information not already incorporated into market prices. Consequently, high accuracy or macro-F1 should not be interpreted automatically as evidence of tradable alpha.

At the same time, the absence of statistically significant daily return predictability does not imply that the sentiment models contain no economically relevant information. The observed disconnect may partly reflect limitations in the downstream dataset, signal construction, and evaluation frequency. Financial news can be incorporated into the prices of liquid equities within minutes or hours. An evaluation beginning in the following trading session may therefore occur after much of the original price response has already taken place. The downstream findings should thus be interpreted as evidence that the evaluated signals did not provide robust \emph{next-session or multi-day} predictability under the present experimental design, rather than as a general rejection of the economic value of financial sentiment.

\subsection{Limitations}

Several limitations qualify the conclusions of this study. First, the
downstream evaluation is restricted to a single calendar year (2019) and a
fixed S\&P~100 universe. The use of 2019 provides a comparatively regular
pre-COVID benchmark period for an initial evaluation of the economic value of
news sentiment. However, this period does not cover major structural
disruptions or a sufficiently broad range of bull, bear, high-volatility, and
low-volatility market regimes. The results therefore may not generalize to
substantially different market conditions. Moreover, although focusing on
large-cap stocks reduces the influence of illiquidity and extreme
microstructure noise, it does not ensure complete or balanced news coverage.
Usable Benzinga headlines are available for only 72 constituents, and the
number of observations is distributed unevenly across firms and trading days.
Frequently covered companies contribute substantially more observations,
while some daily cross-sections contain only a small number of stocks with
fresh news, reducing the stability and statistical power of daily Rank IC
estimates. Future work should extend the evaluation across multiple years,
broader equity universes, and distinct market regimes, while also examining
shorter intraday or event-driven return horizons.

Second, the use of pretrained language models released after the 2019
downstream evaluation period introduces a potential risk of temporal data
contamination. Some Benzinga headlines, or closely related descriptions of the
same events, may have appeared in the models' pretraining corpora. Because the
exact training data are not fully observable, this possibility cannot be
excluded. This is not conventional look-ahead bias in portfolio construction,
since all trading signals and returns are aligned chronologically, but it may
allow the models to use information unavailable to a genuinely point-in-time
2019 system. The downstream results should therefore be interpreted with this
qualification.

Third, the temporal granularity of the experiment may not match the actual price-discovery process of financial news. The study aggregates same-day headlines into a daily stock-level signal and evaluates returns beginning in the following trading session. For liquid large-cap equities, however, public information may be incorporated into prices within minutes or hours of publication. As a result, the experiment may miss economically meaningful intraday reactions and instead measure the residual effect after the primary adjustment has occurred. This timing mismatch offers a plausible explanation for the small one-day ICs and their subsequent decay.

Fourth, daily aggregation removes potentially important event-level information. Multiple headlines concerning the same company may differ in novelty, importance, reliability, and relevance. Repeated or syndicated headlines may receive excessive weight, while simultaneous positive and negative events may offset one another in the aggregated score. The current design also does not explicitly distinguish among scheduled announcements, analyst actions, earnings news, legal events, and general corporate reporting. Moreover, probability outputs from different model families are not necessarily calibrated on a common scale, which limits direct comparisons between probability-weighted sentiment signals.

Fifth, the model objective itself is not directly aligned with asset-return prediction. The models were optimized to reproduce human sentiment annotations rather than forecast abnormal returns. A headline can be correctly classified as positive even when the information is anticipated, already priced, immaterial relative to expectations, or associated with a negative market reaction. Financial price responses depend not only on semantic polarity but also on surprise, novelty, investor expectations, market conditions, and prior positioning. This objective mismatch limits the extent to which improvements in classification performance can be expected to translate mechanically into higher IC or portfolio returns.

Sixth, the reported portfolio results are gross and exclude commissions, bid--ask spreads, market impact, slippage, and short-borrowing costs. These frictions would be particularly important for a high-turnover news-based strategy. The portfolios were also not explicitly neutralized against market beta, industry, firm size, momentum, or other established risk factors. Long-only returns therefore cannot be interpreted as pure sentiment alpha, while the long--short portfolios may retain residual systematic exposures. The strategies should be regarded as controlled signal comparisons rather than directly implementable investment products.

Finally, this study evaluates a limited selection of model architectures, QLoRA configurations, prompts, and hyperparameters. Alternative rank choices, probability calibration procedures, class-weighted objectives, longer textual inputs, or return-aware training objectives could generate different results. The observational design also identifies associations rather than establishing a causal relationship between news sentiment and subsequent stock prices.

\subsection{Future Research}

Future work should first expand the downstream dataset across a larger equity universe, a longer historical period, and multiple market regimes. A broader point-in-time news sample would increase the number of active-news stocks in each daily cross-section and permit separate analysis by firm size, sector, liquidity, news frequency, and market condition. Controlling for unequal ticker coverage and repeated news would further reduce potential selection and concentration biases.

A particularly important extension is intraday event-based evaluation. Timestamped headlines can be aligned with prices immediately before publication and assessed over horizons such as 5 minutes, 30 minutes, 1 hour, and 4 hours:

$$
R_{i,t}^{(h)} =
\frac{P_{i,t+h}}{P_{i,t^-}}-1,
\qquad
h\in\{5\text{ min},30\text{ min},1\text{ hour},4\text{ hours}\},
$$

where $P_{i,t^-}$ denotes the price immediately before the news event. Separating pre-market, intraday, and after-market announcements would provide a more precise match between news arrival and price discovery. Market- or sector-adjusted abnormal returns could then be calculated as

$$
AR_{i,t}^{(h)}=R_{i,t}^{(h)}-R^{(h)}_{\text{benchmark},t}
$$

thereby reducing contamination from broad market movements.

Further research should also incorporate event novelty, source reliability, article importance, earnings surprises, analyst revisions, and the interaction between sentiment and investor expectations. Probability calibration and continuous sentiment scores may preserve more cross-sectional information than discrete labels. Finally, models could be trained with multi-task objectives that combine sentiment classification with return direction, abnormal-return magnitude, or event-impact prediction, bringing the training objective closer to the intended financial application.

In conclusion, this study shows that QLoRA can adapt large language models effectively and efficiently to financial sentiment classification, but it does not find statistically robust evidence that the resulting daily sentiment signals predict subsequent stock returns. The downstream results identify an important gap between linguistic accuracy and financial usefulness, while also showing that this gap may be amplified by limited ticker coverage, uneven news frequency, short sample duration, daily aggregation, and a mismatch between news-arrival timing and the evaluation horizon. A definitive assessment of the economic value of LLM-based sentiment therefore requires broader point-in-time data, intraday evaluation, risk-adjusted portfolio construction, and objectives designed more directly around market reactions.

\section*{Ethics and Privacy Statement}

This study uses financial text and market data for research evaluation and does
not involve human-subject intervention or private personal information.
Model-generated sentiment and backtest results may be erroneous or unstable and
should not be interpreted as individualized investment advice. Reproducibility
also depends on respecting the licenses and usage conditions of the underlying
datasets and pretrained models.


\end{document}